\documentclass{webofc}
\usepackage[varg]{txfonts}   % Web of Conferences font
\usepackage{multirow} 
\usepackage{amsmath} 
\usepackage{amssymb}
\usepackage{epsfig} 
\usepackage{graphicx} 
\usepackage{color}
\usepackage{xspace} 
\usepackage{lineno}
\usepackage[varg]{txfonts}   % Web of Conferences font
\usepackage{subfigure}
\usepackage{graphicx}
\newcommand{\AuAu}{\mbox{${\rm Au+Au}$}\xspace}

\newcommand{\pp}{\mbox{${p+p}$}\xspace}

\newcommand{\mgg}{\mbox{${\rm M_{\gamma\gamma}}$}\xspace}

\newcommand{\snn}{\mbox{$\sqrt{s_{_{NN}}}$}\xspace}
\newcommand{\s}{\mbox{$\sqrt{s}$}\xspace}

\newcommand{\pt}{\mbox{$p_{T}$}\xspace}

\newcommand{\piz}{\mbox{$\pi^{0}$}\xspace}

\newcommand{\vtwo}{\mbox{${\rm v_{_{2}}}$}\xspace}

\usepackage{xspace}	% Include xspace

\newcommand{\pio}{\mbox{$\pi^0$}\xspace}

\newcommand{\ppm}{\mbox{$p^\pm$}\xspace}

\newcommand{\pimp}{\mbox{$\pi^\mp$}\xspace}

\def\detdeta{\mbox{$dE_T/d\eta$}\xspace}

\newcommand{\Ks}{\mbox{$K_{S}^{0}$}\xspace}			
\newcommand{\mKsdecay}{\mbox{$\pi^{+}\pi^{-}$}\xspace}

\newcommand{\Lambdaz}{\mbox{$\Lambda^{0}$}\xspace}

\newcommand{\vpi}{$v_2^{\pi^0}$}
\def\FigureOne{
\begin{figure*}[!]
\vskip -1cm
\centering
\includegraphics[width=0.95\linewidth]{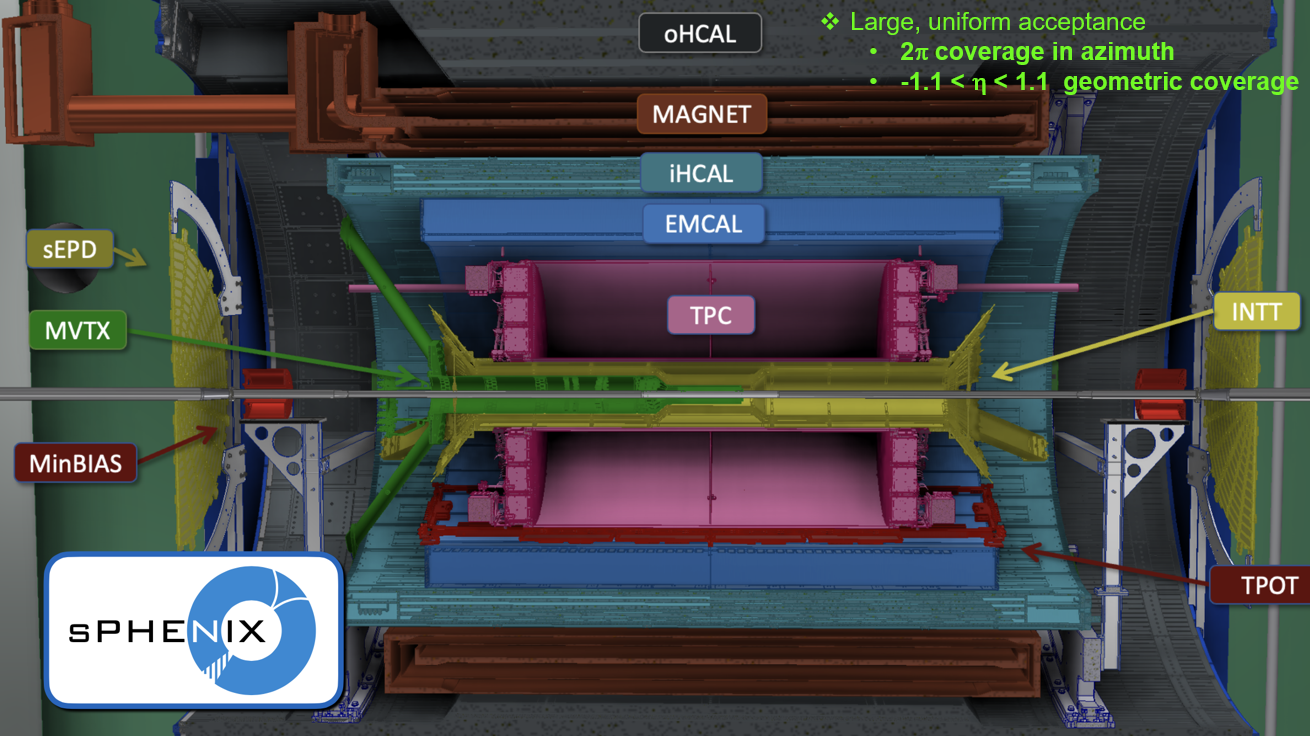}
\vskip -0.2cm
\caption{A Schematic of side view of the sPHENIX detector during RHIC 2023 and 2024 Runs.}
\label{fig:fig1}
\end{figure*}
}

\def\FigureTwo{
\begin{figure*}[!bht]
\vskip -0.5cm	
\subfigure[\Ks peak measurement in \pp at 200 GeV]
{	
\includegraphics[width=0.5\linewidth]{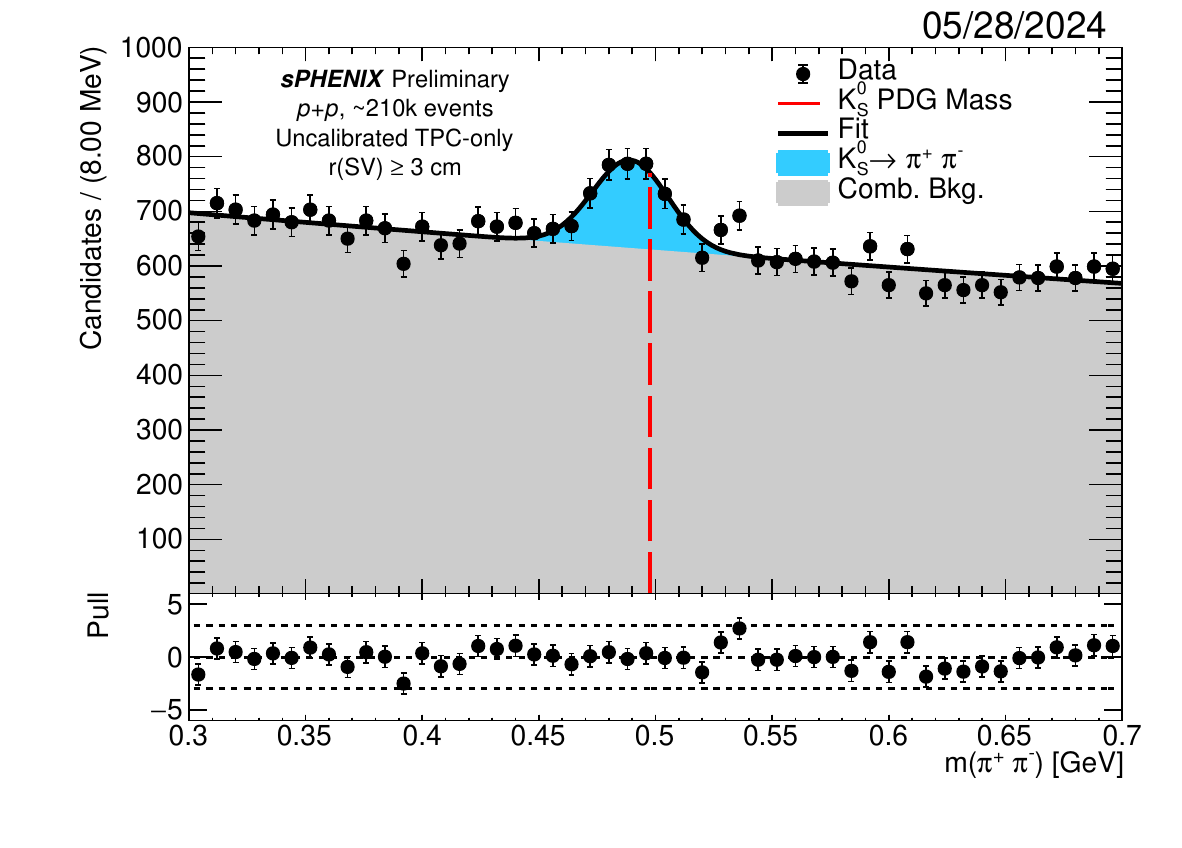}
\label{fig:fig2a}	
}
\subfigure[\Lambdaz peak measurement in  \pp at 200 GeV]
{
\hspace{-0.5cm} 
\includegraphics[width=0.5\linewidth]{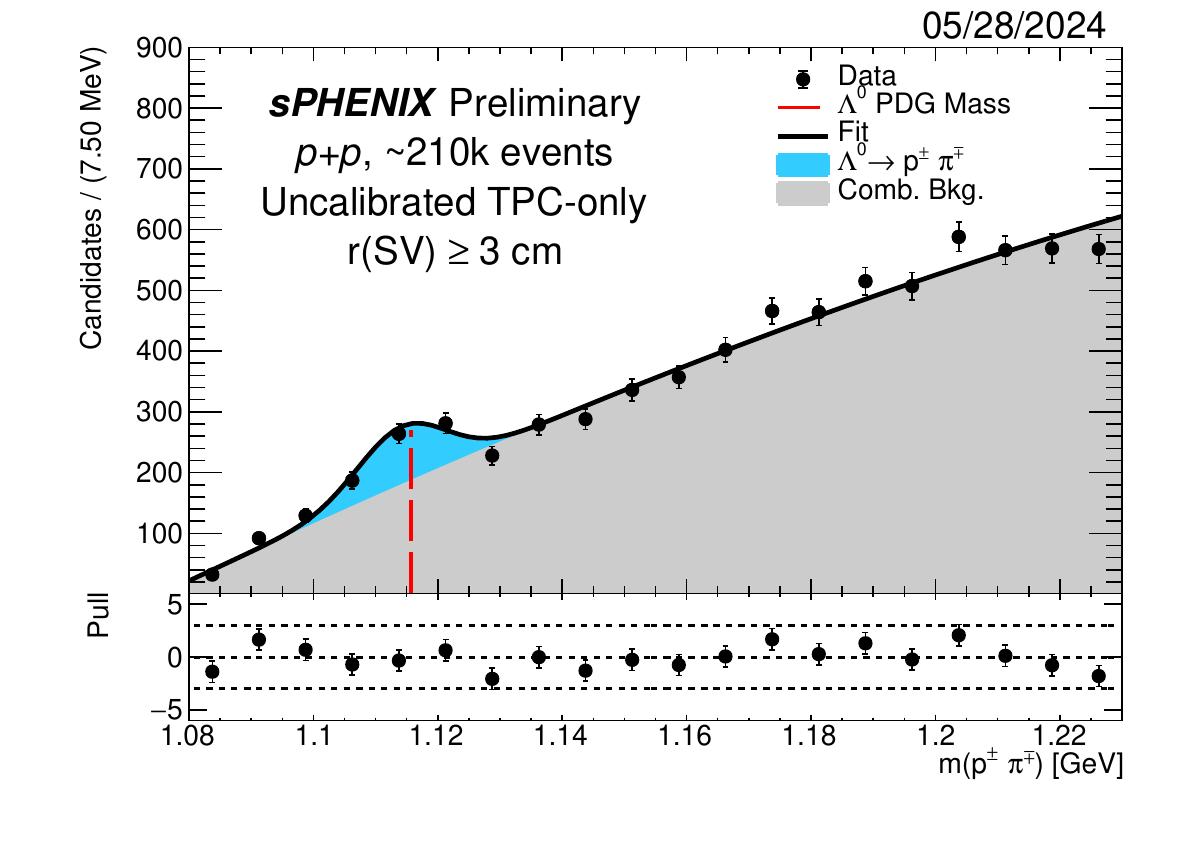}	
\label{fig:fig2b}
}
\vskip -0.2cm
\caption{Measured invariant mass spectra by the sPHENIX-TPC detector: (a) \mKsdecay, and (b) \ppm \pimp obtained in \pp collisions at \s~=~200~GeV from RHIC 2024-Run. The combinatorial background and signals are shown in gray and blue, respectively.}     	
\label{fig:fig2}
\end{figure*}
}

\def\FigureThree{
	\begin{figure*}[!hbt]
		\vskip -1cm
		\centering
		\includegraphics[width=1.0\linewidth]{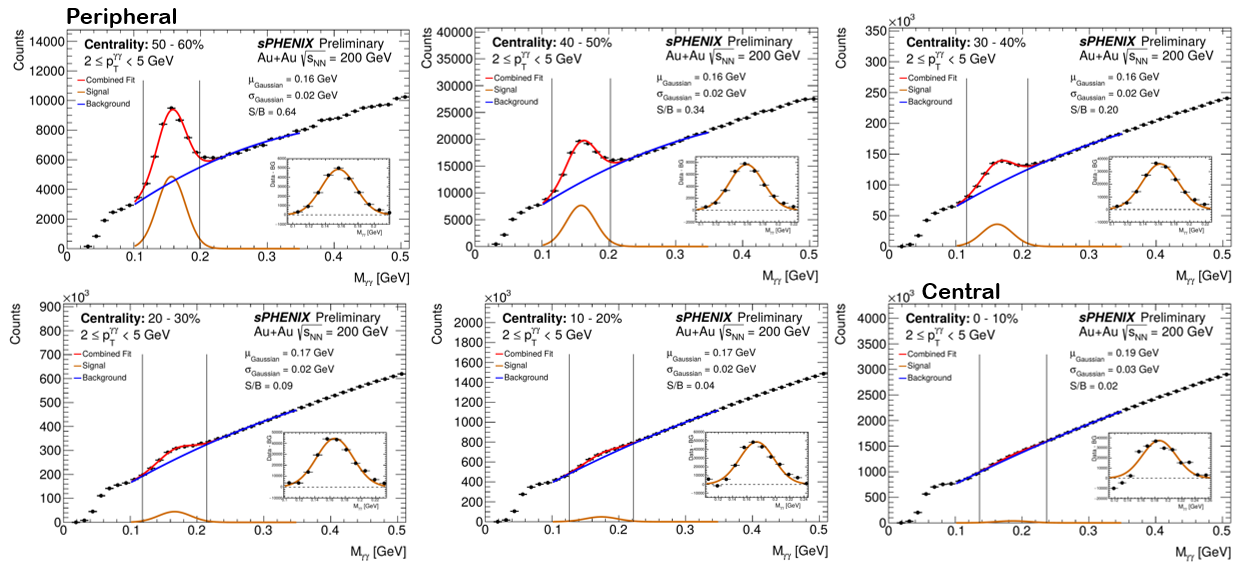}
		\vskip -0.4cm
		\caption{Diphoton invariant mass \mgg spectrum for each centrality bin measured by sPHENIX in \AuAu at \snn = 200 GeV during detector commissioning in the RHIC 2023 Run. In each panel, the red curve is the total fit, the orange curve is the Gaussian used to fit the signal, and the blue curve is the polynomial used to fit the combinatorial background. The insets show the fitted mass peak after background subtraction. The uncertainties in these invariant mass distributions are statistical only.}
		\label{fig:fig3}
	\end{figure*}
}

\def\FigureFour{
	\begin{figure*}[!]
		\vskip -0.5cm
		\centering
		\includegraphics[width=0.6\linewidth]{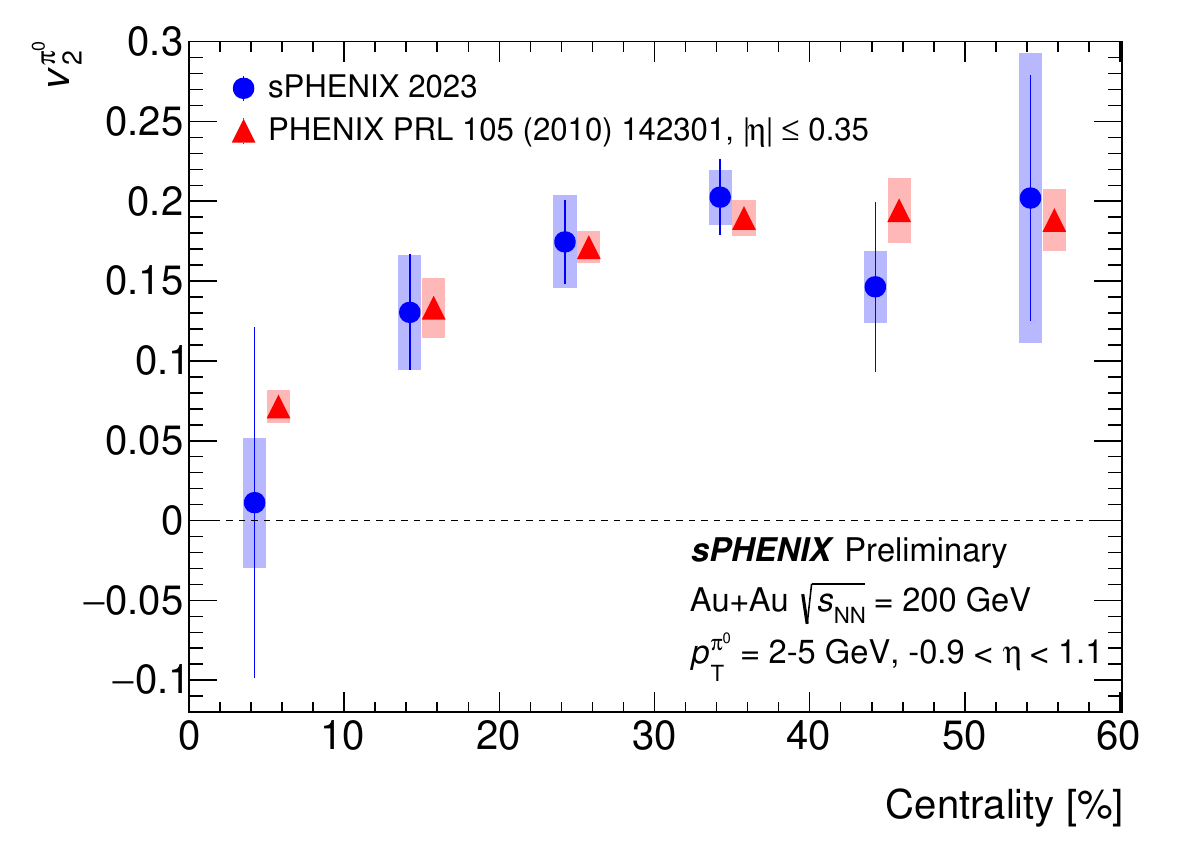}
		\vskip -0.5cm
		\caption{Elliptic flow $v_2^{\pi^0}$ as a function of centrality measured by sPHENIX in \AuAu at \snn~=~200~GeV during detector commissioning in the RHIC 2023 Run. These measurements are integrated over the range of $2\le p_\text{T}\le5$ GeV. The results, shown in blue with both statistical and systematic uncertainties, are offset to the left, while PHENIX data (integrated over the same \pt range as the sPHENIX points)~\cite{phenixpi0flow}, shown in red, is displaced to the right in each centrality bin for visibility.}
		\label{fig:fig4}
	\end{figure*}
}

\def\FigureFive{
	\begin{figure*}[!hbt]
		\vskip -0.5cm	
		\subfigure[\detdeta measurements EmCal vs HCal]
		{	
			\includegraphics[width=0.5\linewidth]{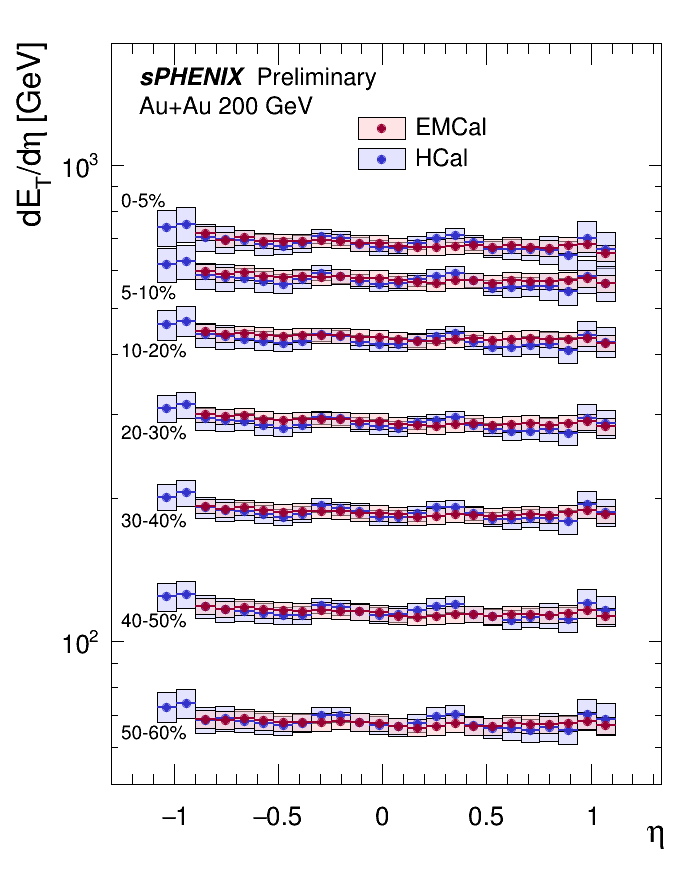}
			\label{fig:fig5a}		
		}
		\subfigure[\detdeta measurements from sPHENIX, PHENIX, and STAR]
		{
			\hspace{-0.5cm} 
			\includegraphics[width=0.5\linewidth]{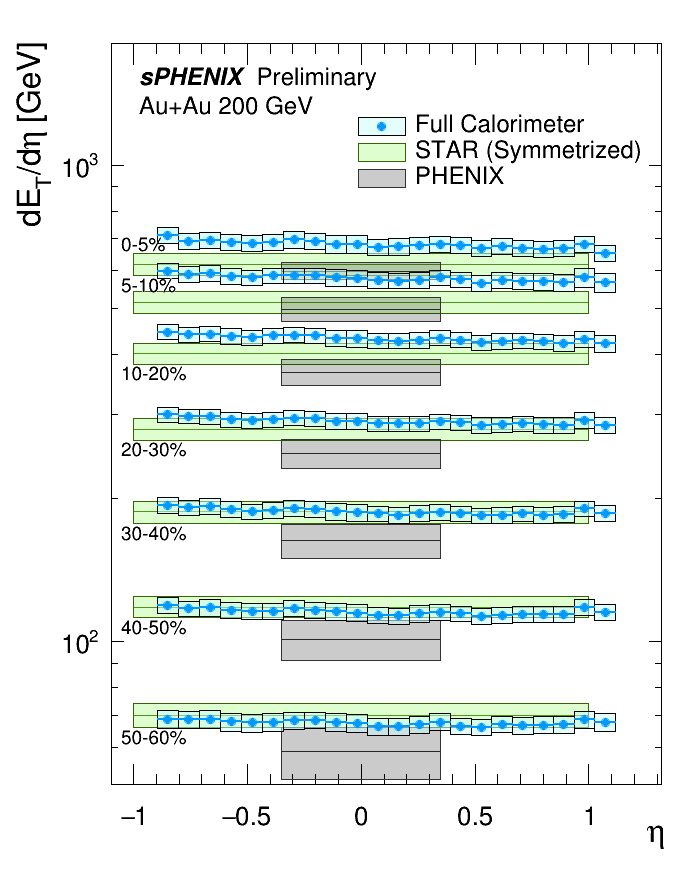}
			\label{fig:fig5b}
		}
			\caption{Measured \detdeta for different centrality bins obtained in \AuAu at \snn~= 200 GeV using detector commissioning data from the RHIC 2023 Run. (a) comparison of \detdeta measurements for EmCal-only results ($-0.9 < \eta < 1.1$) and HCal-only results ($-1.1 < \eta < 1.1$). (b) Comparison of \detdeta measurements for different centrality bins obtained by sPHENIX, PHENIX, and STAR experiments.}  
		\vskip -0.2cm 	
		\label{fig:fig5}
	\end{figure*}
}

\begin{document}
\vglue -2cm
\title{sPHENIX Highlights: First Results from sPHENIX at RHIC}
\makeatother  
\author{\firstname{Rachid} \lastname{Nouicer}
\inst{1}\fnsep\thanks{\email{nouicer@bnl.gov}}\ (for the sPHENIX Collaboration) } \institute{Physics Department, Brookhaven National Laboratory, Upton, New York 11973, United States}
\abstract{
First results from the sPHENIX experiment on the \pio~\vtwo and \detdeta in Au+Au collisions at \snn = 200 GeV using detector commissioning data during the RHIC 2023 Run are presented. These results are shown across a large centrality range, and compared to previous PHENIX and STAR results. These measurements demonstrate the sPHENIX capabilities towards accomplishing the sPHENIX jet physics program. From the ongoing RHIC 2024 Run, \pp collisions at \s = 200 GeV, a first glimpse of the measurements \mbox{\Ks and \Lambdaz} obtained by the sPHENIX Time Projection Chamber (TPC) detector are shown.
}

\maketitle
\vspace*{- 0.5cm}
\section{Introduction \label{sec-1}}
 sPHENIX is a new experiment currently operating at the Relativistic Heavy Ion Collider (RHIC) at Brookhaven National Laboratory (BNL). The sPHENIX detector was commissioned and took first \AuAu data during the RHIC 2023 Run~\cite{sPHENIX,QM2023OBrien}. The sPHENIX experiment provides state-of-art capabilities for studies of the strongly interacting quark-gluon plasma using jet and heavy-flavor observables. The goal of sPHENIX is to understand the microscopic structure of the plasma and reveal how its strongly interacting nature arises from the underlying interactions of quarks and gluons described by quantum chromodynamics~\cite{sPHENIXPhy}.
 
 This paper presents first sPHENIX measurements on the \pio \vtwo and \detdeta in Au+Au collisions at \snn = 200 GeV obtained during detector commissioning in the \mbox{RHIC 2023 Run}. In addition, a first look at the \mbox{\Ks and \Lambdaz} measured by the sPHENIX TPC in \pp collisions at \s = 200 GeV from the ongoing RHIC 2024 Run are presented.
\section{sPHENIX Detector\label{sec-2}}
The sPHENIX detector is a cylindrical spectrometer covering $\mid\eta\mid \leq$ 1.1 and the full azimuth. Figure~\ref{fig:fig1} shows a schematic of the sPHENIX detector components during RHIC 2023  and 2024 Runs. It uses the former 1.4 Tesla BaBar superconducting solenoid magnet to contain an inner tracking system followed by electromagnetic and hadronic calorimeters. The sPHENIX subdetectors are as follows.
\begin{itemize}
\item [\bf 1)] \noindent{\bf\ Tracking system}: the tracking system from inside-out radially consists  of three components:
\begin{itemize}
{\bf \item MAPS Vertex Tracker (MVTX):} the MVTX sensors are a Monolithic Active Pixel Sensors with pixel pitch of 27 $\mu$m. It is a copy of the inner three layers of the ALICE Inner Tracking System (ITS) with layers at radial positions of 2.5, 3.2, and 4.0~cm. The MVTX three-layer silicon pixel detector are located just outside the sPHENIX beryllium beam-pipe. The MVTX measures the collision vertex, the displaced vertex of charged tracks, and contributes to the overall tracking resolution of sPHENIX.
{\bf \item Silicon Strip INTermediate Tracker (INTT)}: 
the INTT is a silicon strip detector with strip pitch 78 $\mu$m in $\phi$, consisting of two hermetic cylindrical barrels.  The INTT plays key roles in the sPHENIX experiment; 1) it provides seeds and enhances the resolution of the track reconstruction of the charged particles, and 2) it has 60 ns time resolution allowing it to read out collision data from each single RHIC  beam bunch crossing and suppress event pileup background. 
{\bf \item Time Projection Chamber (TPC)}: 
the TPC is a compact cylindrical tracking detector, 2.1 m in length with
an outer radius of 80 cm and a total gas volume of $\approx$ 4 m$^3$ . The TPC operates
with an Ar-CF$_{4}$-Isobutane 75/20/5 gas mixture. The TPC field cage is constructed from carbon fiber/honeycomb for mechanical strength and use 18 layers of 3~mil kapton to insulate high voltage (45 kV) from ground. The device is read out on each end with quad-GEM modules each containing four layers of GEM foils plus a cathode pad plane. There are 36 quad-GEM modules on each end of the TPC. The TPC makes high precision measurements (space point resolution < 200 $\mu$m) of the momenta of all charged tracks passing through sPHENIX at mid-rapidity. Figure~\ref{fig:fig1} shows a first look at the \Ks and \Lambdaz peaks 
in \pp collisions at \s = 200 GeV from the ongoing RHIC 2024 Run. These measurements were obtained using only the TPC reconstructed tracks passing a loose selection cuts with no corrections to the measured track trajectory. 
\FigureOne
\FigureTwo
{\bf \item Time Projection Outer Tracker (TPOT)}:
the TPOT consists of eight identical Micromegas modules, grouped in three sectors~\cite{TPOT}. The three sectors are mounted to the Electromagnetic Calorimeter (EmCal) at the bottom facing TPC. Each TPOT module measures a space point of charged tracks passing through TPC. Its function is to provide tracking distortion correction-information for the TPC.
\end{itemize}
\item [\bf 2)] \noindent{\bf Electromagnetic Calorimeter (EmCal)}: 
just outside the TPC/TPOT, the EmCal detector is an electromagnetic calorimeter with solid angle coverage of $|\eta|<1.1$ and $2\pi$ in $\phi$. The EmCal is 20.1 radiation lengths deep and is designed to measure photons, electrons and positrons via electromagnetic showers. The sPHENIX EmCal is constructed from a tungsten powder absorber with embedded scintillating fibers to make towers of size ${\Delta \eta \times \Delta \phi =}$ 0.024 $\times$ 0.024. The light from the scintillating fibers is collected by a light guide onto four gain-matched silicon photomultipliers (SiPMs). The EmCal is 0.83 hadronic interaction lengths deep and therefore sees a significant amount of hadronic shower energy~\cite{EmCal}.
\item [\bf 3)] \noindent{\bf Hadronic Calorimeter (HCal)}: 
the HCal consists of two components; Inner Hadronic Calorimeter (IHCal) located inside the magnet cryostat (just after EmCal), and Outer Hadronic Calorimeter (OHCal) located outside the magnet. Both IHCal and OHCal are cylindrical. The HCal is 4.9 hadronic interaction lengths thick. Both the IHCal and OHCal are sampling calorimeters consisting of aluminum (inner)/steel (outer) absorbing plates and scintillating tiles with tower size ${\Delta \eta \times \Delta \phi =}$ 0.1 $\times$ 0.1. Both IHCal and OHCal scintillating tiles are set at an angle offset to the transverse direction to reduce channeling particles that do not interact with the active volumes of these calorimeters. The light from these tiles is collected into SiPMs. 
\item [\bf 4)] \noindent{\bf Global Detectors} are comprised of following elements: 
\begin{itemize}
{\bf \item Minimum Bias Trigger Detector (MBD)} consists of 64 photomultiplier arrays with quartz radiator (PMTs) on each of the two arms covering the very forward region in $3.61<|\eta|<4.51$. The PMTs are arranged in 3 concentric rings around the beam-pipe, covering $2\pi$ in azimuth. The MBD system timing resolution is 50~ps. The MBD is used for minimum bias, centrality-selected event triggering.
{\bf \item sPHENIX Event Plane Detection (sEPD)}: the sEPD is a pair of disks of scintillating tiles finely segmented in $\phi$ with readout into SiPMs located at forward pseudorapidity $2 <|\eta|< 4.9$. Its role is high-resolution event plane determination.
{\bf \item Zero Degree Calorimeters (ZDC)}: 
the ZDC are small transverse area hadron calorimeters located downstream of the DX dipole magnets in each side of the sPHENIX experiment about 18 m from the interaction point. It is interleaved layers of tungsten and PMMA optical fibers readout with photomultiplier tubes. The ZDC measure neutral energy within a 2 mrad cone about the beam direction. 
\end{itemize}
\end{itemize}
\FigureThree
\FigureFour
\section{First Results\label{sec-3}}
\subsection{\small Measurement \pio \vtwo in \AuAu Collisions at \snn~=~200 GeV From RHIC 2023 Run \label{subsec-3a}}
The data used for this analysis was taken with the EmCal, and MBD detectors. 
The \mbox{EmCal} is used to measure photons to reconstruct \piz's, and the MBD is used for minimum bias, centrality-selected event triggering, and event centrality determination~\cite{Emma}.  In addition, signals from the MBD PMTs are used within this analysis to calculate the flow reaction plane in collision events. During the 2023 commissioning run, subsystems were in partial operation, and the EmCal was read out in the region $-0.9 < \eta < 1.1$. The EmCal data analysis starts by doing a tower-by-tower energy calibration and making the tower response uniform in $\phi$. 
Pairs of clusters are then found and the invariant mass of the pair is determined. $\eta$-dependent calibration constants which move the $\pi^0$ mass peak to the expected value are extracted based on the $\eta$ location of the most energetic tower within the higher energy cluster of the pair. This procedure is done iteratively until the calibration is stable for all $\eta$ rings. In reconstructing the $\pi^0$ candidates, each pair of clusters in the event that satisfies cuts of $\chi^2 < 4$ and \mbox{$E_{\text{cluster}} \ge 1$ GeV} is considered. The \mbox{$E_{\text{cluster}} \ge 1$ GeV} requirement is applied to exclude clusters that originate from noise in the EmCal. After applying the selection criteria, invariant mass ($M_{\gamma\gamma}$) distributions of \piz candidates with \mbox{$2 \le p_\text{T} \le 5$ GeV} are generated for six centrality selections from $0\mbox{--}60\%$. The distributions are fit with a combined Gaussian and second-order polynomial function.

Figure \ref{fig:fig3} shows the diphoton invariant mass \mgg spectrum for each centrality bin measured by sPHENIX in \AuAu at \snn = 200 GeV during detector commissioning in the RHIC 2023 Run. The vertical black lines mark the signal bounds, calculated in the range $\mu_{\text{Gauss}} \pm 2\sigma_{\text{Gauss}}$. Here, $\mu_{\text{Gauss}}$ denotes the Gaussian mean (peak position), and $\sigma_{\text{Gauss}}$ the standard deviation. The fit is performed over the range  $0.1<M_{\gamma\gamma} <0.35$ GeV, with the upper bound optimized to adequately cover the background while excluding the influence of the $\eta$ meson (with \mbox{$M_{\gamma\gamma}=0.55$ GeV}), and the lower bound is set to exclude the region \mbox{$M_{\gamma\gamma} < 0.1$ GeV}~\cite{Emma}. 

Figure \ref{fig:fig4} shows the results of elliptic flow, \vpi, integrated over the range of $2\le p_\text{T}\le5$ GeV as a function of centrality bins. The results are overlaid with $p_\text{T}$-integrated data from PHENIX measurement~\cite{phenixpi0flow}. Good agreement is observed between the two measurements across the full centrality range studied. 
\subsection{\small Measurement of \detdeta in \AuAu Collisions at \snn~=~200 GeV From  RHIC 2023 Run \label{subsec-3b}}
To obtain \detdeta measurements, we use data from all three calorimeters, EmCal, IHCal, and  OHCal. In addition, the MBD detector is used for determination of the collision vertex and the centrality of the event triggering. 
\FigureFive

In the present analysis~\cite{Emma}, the absolute energy scale of the EmCal is established from an $\eta$-dependent calibration of the $\pi^{0}$ meson peak in data to the same position as in simulation. The $\eta$ region for the EmCal is limited to $-0.9 < \eta < 1.1$ because the EmCal readout electronics were only partially instrumented during this stage of commissioning. It should be noted that the calibration of the EmCal is the same as in the \pio \vtwo  analysis presented in the previous section.
The IHCal and OHCal absolute EM energy scale calibration is performed using the minimum ionizing particle energy depositions from cosmic ray muons from data taken in early 2024. Temperature-dependent corrections to the detector energy scale are applied to these cosmics calibrations to account for gain variations from detector conditions during collision data-taking. The uncorrected \detdeta is calculated for a given centrality class as the sum of calorimeter tower $E_{\text{T}}$ as a function of $\eta$. The correction factors for each calorimeter sub-system are fairly constant with centrality.
%,with about 66\% of total \detdeta reconstructed by the EMCal, about 14\% reconstructed by the OHCal and about 4\% of the $E_{T}$ reconstructed by the IHCal in the calorimeter $\eta$ acceptance with both the IHCal and OHCal calibrated to the EM scale. 

In Fig.~\ref{fig:fig5a}, the EmCal-only and HCal-only \detdeta measurements are overlaid to highlight their agreement. This is a particularly encouraging result as the EmCal and HCal see different contributions of the collision energy. Further, for all calorimeter measurements, \detdeta at positive $\eta$ and negative $\eta$ are compatible within uncertainties. 
The uncertainty contributions to the measurement include the calorimeter energy response, the application of the particle spectra reweighing in simulation, calorimeter noise, detector acceptance, and $z$-vertex resolution effects.

Figure~\ref{fig:fig5b} shows comparison of the present \detdeta measurements from sPHENIX to published results from the PHENIX \cite{PHENIX_detdeta} and STAR \cite{STAR_detdeta} experiments. The sPHENIX results are consistently higher than the results from PHENIX for all centrality bins but agree within uncertainties for mid-central bins 30-60\%; the sPHENIX results are above the STAR results in the centrality range of 0-10\% but are in agreement in other centrality intervals. Presently, these sPHENIX results use a preliminary centrality calculation for Run 2023 data, which is expected to be updated before publication.

\section{Summary \label{sec-4}} 
sPHENIX measured \pio \vtwo and \detdeta in \AuAu at \snn~=~ 200 GeV obtained during detector commissioning in the RHIC 2023 Run. These results are measured with the calorimeter system, including the hadronic calorimeter, which is used for the first time for such observables at RHIC. Results are shown as a function of $\eta$ across a large centrality range, and compared to previous PHENIX and STAR results demonstrating good understanding of the sPHENIX calorimeter response, and laying the path towards accomplishing the sPHENIX jet and heavy flavor physics program at RHIC.

\end{document}